\documentclass[conference]{IEEEtran}

  	\usepackage[pdftex]{graphicx}
  	\graphicspath{{../pdf/}{../jpeg/}}
	\DeclareGraphicsExtensions{.pdf,.jpeg,.png}

	\usepackage[cmex10]{amsmath}
	\usepackage{mathabx}
	\usepackage{algorithmic}
	\usepackage{array}
	\usepackage{mdwmath}
	\usepackage{mdwtab}
	\usepackage{eqparbox}
    \usepackage{authblk}
	\usepackage{url}
    \usepackage{xcolor}
    \usepackage[dvipsnames]{xcolor}
\title{\LARGE Impact of Pump Phase-Noise on Josephson Traveling-Wave Parametric Amplifiers}

\author[1,$\dagger$]{Daryoush Shiri}
\author[1]{Likai Yang}
\author[1]{Saesun Kim}
\author[2]{Mohamed A. Hassan}
\author[1]{Philip Krantz}
\author[1]{Eric T. Holland}

\affil[1]{Quantum Engineering Solutions (QES), Keysight Technologies Inc.,
Santa Rosa, CA 95403, USA}
\affil[2]{Design Engineering Software (DES), Keysight Technologies Inc., 
Santa Rosa, CA 95403, USA}
\affil[$\dagger$]{daryoush.shiri@keysight.com}

\begin{document}
\maketitle

\begin{abstract}
Superconducting traveling-wave parametric amplifiers (TWPAs) are essential elements for enhancing the signal-to-noise ratio (SNR) and thus the read-out fidelity of superconducting qubits because of their high gain and near quantum-limited noise. However, the impact of the pump source, \textit{e.g.,} phase noise on these amplifiers, has not yet been studied. In this work, we show that among the two amplification processes in JTWPAs, the three-wave mixing (3WM) process is more sensitive to the pump phase noise than the four-wave mixing (4WM) process. We show that the even-order nonlinearity of 4th order and above in three-wave mixing is responsible for more than 10 dB increase of phase noise at high frequency offsets within the phase noise mask as the power of the pump increases. A polynomial model of the amplifier and cyclo-stationary property of phase noise also corroborate with the simulations. The Harmonic Balance (HB) periodic noise analysis tool and Leeson phase noise model in Keysight Advanced Design System (ADS) simulator were used in this study.
\end{abstract}

\IEEEoverridecommandlockouts

\IEEEpeerreviewmaketitle

\section{Josephson Traveling-Wave Parametric Amplifier}
\begin{figure}[t!] 
\centering
\includegraphics[width=3.5in]{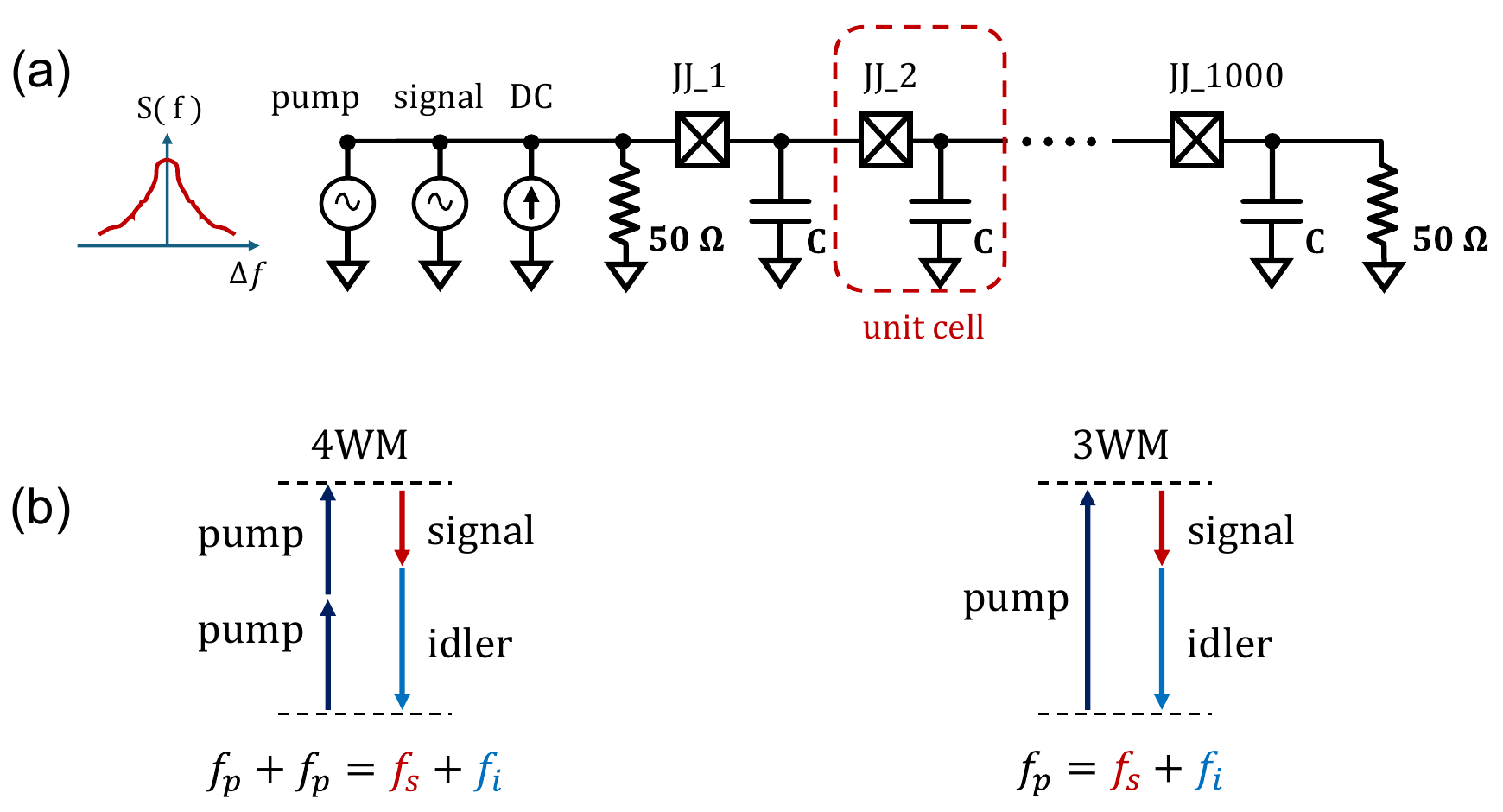}
\caption{(a) Schematic of a JTWPA with 1000 unit cells made of a JJ (nonlinear inductor) and a capacitor. The pump source in this study has phase noise (the left-most input source). (b) The four-wave mixing (4WM) and three-wave mixing (3WM) processes involving down-conversion of pump photon(s) ($f_p$) to a signal ($f_s$) and an idler photon ($f_i$).}
\label{1000jj}
\end{figure}

A traveling-wave parametric amplifier based on superconducting elements is in essence, implementation of a nonlinear transmission line (NTL) \cite{Landauer1960}. The nonlinear inductance of the NTL emanates from the physics of the Josephson junction (JJ) \cite{Josephson1962, krantz2019quantum,shiri2024modeling,aumentado2020superconducting} or the kinetic inductance of the superconductor, which is also a nonlinear function of the electric current \cite{Mohebbi09,malnou2021three}. The distributed amplification takes place by injection of a high-power pump which modulates the nonlinear Josephson inductance along the NTL. As a result of frequency mixing, the pump power is transferred to the signal and the idler, and under the right conditions, amplification of the signal will ensue as shown in Fig. \ref{1000jj}. The mathematical model for JJ was implemented in Keysight ADS in its entirety, without any approximations like Taylor expansion. The nonlinearity of Josephson inductance originates from its dependency on the current, $I$, passing through the JJ and it is found from:

\begin{equation}
\label{Ljj}
L_{J}(I)=\frac{L_{J0}}{\sqrt{1-\left(\frac{I}{I_c}\right)^2}},
\end{equation}

\noindent where $I_c$ is the critical current of the JJ and $L_{J0}=\Phi_0/(2\pi I_c)$ is the zero-bias inductance, where $\Phi_0=h/2e \approx 2.07\times 10^{-15}~\text{Wb}$ is the magnetic flux quantum. For our simulation, we set the critical current of the JJ to $1.4~\mu$A --- representative of the value reported in previous experiments \cite{malnou2021three,Fadavi2023,macklin2015near}.

\begin{figure}[t!] 
\centering
\includegraphics[width=3.5in]{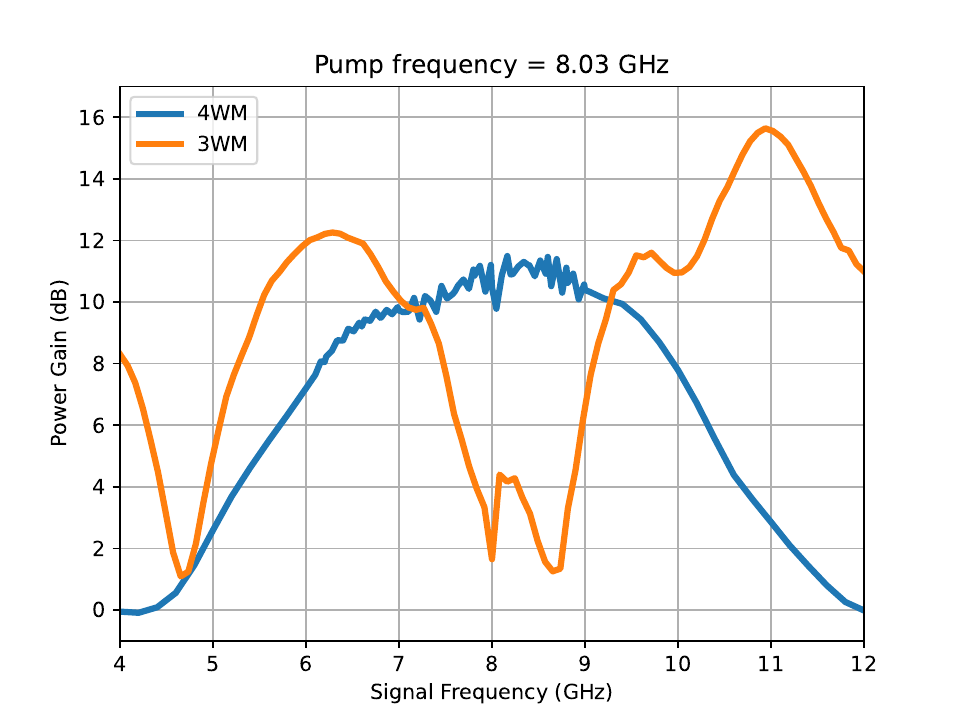}
\caption{The power gain of 4WM JTWPA (zero DC bias) versus the signal frequency (blue). The pump current amplitude is $I_p = 1.4~\mu A$. The power gain of 3WM JTWPA (DC bias $I=0.5I_c$) versus the signal frequency (orange). The pump current amplitude is $I_p = 0.6~\mu\text{A}$. For both cases the pump frequency is $f_p=8.03~\text{GHz}$ and the input signal power is $P_{s}=-150~\text{dBm}$.}
\label{gain_both}
\end{figure}

The capacitance ($C$) and inductance ($L_{J0}$) of each unit cell are designed such that the characteristic impedance of the transmission line \textit{i.e.,} $Z_c\approx\sqrt{L/C}$, is $50~\Omega$ at given bias current. With $C = 93~\text{fF}$ the cut-off frequency of the TL is found from the dispersion of the TL in the linear regime and it is $f_{\text{cutoff}}=1/\pi\sqrt{LC}= 63 ~\text{GHz}$ (at $I=0$). If the DC bias current of the JTWPA is zero, the nonlinearity of the JJ inductance creates only odd harmonics in response to the large modulating pump. This leads to the four-wave mixing (4WM) process \cite{macklin2015near,Likai2025} [see Fig. \ref{1000jj}(b)]. This process involves intermixing of four microwave photons \textit{i.e.,} the energy of two pump photons is converted to one signal and one idler photon \textit{i.e.,} $f_p + f_p = f_s + f_i$. This process is akin to the $\chi^{(3)}$ process in nonlinear optics \cite{Armstrong1962}. The conservation of momentum also mandates the equality of photon momenta or wave vectors \textit{i.e.,} $k_p + k_p = k_s + k_i$. This is the phase matching condition for signal amplification. The power gain of the JTWPA for zero DC bias is plotted versus the input signal frequency in Fig. \ref{gain_both} (blue plot). As the physics of 4WM process mandates \cite{macklin2015near}, the gain is symmetric around the pump frequency (here 8.03~GHz). The signal power is fixed at $-150~\text{dBm}$. If the DC bias current is set to $I=0.5I_c=0.7~\mu\text{A}$, the gain mechanism is three-wave mixing (3WM) one \cite{HamnpusThesis}. In this case, the JJ nonlinear inductance at non-zero DC bias creates all harmonics of the pump including the even ones. This effect is akin to the $\chi^{(2)}$ process in nonlinear optics in which one pump photon is down-converted to a signal and an idler photon, \textit{i.e.,} $f_p = f_s + f_i$ [See Fig. \ref{1000jj}(b)] \cite{Armstrong1962}. The gain spectrum for 3WM process is shown in Fig. \ref{gain_both} (orange plot) for $f_p = 8.03$~GHz and the pump current amplitude of $I_p=0.6~\mu\text{A}$. The signal power at the JTWPA input is -150~dBm. 
The gain is above 10~dB for a bandwidth of more than 1~GHz. The extra gain above the pump frequency emanates from other mixing processes with higher pump harmonics \cite{HamnpusThesis}. To avoid those processes and achieve a flatter gain profile, JTWPAs are usually designed with dispersion engineering \cite{macklin2015near}, non-uniform distribution of critical currents \cite{Peng22floquet}, band-pass filtering of higher pump harmonics \cite{Fadavi2023}, or apodizing \textit{i.e.,} spatial modulation of NTL impedance \cite{Sandbo2025}. 
The aim of this study however, is not to enhance the gain or bandwidth, and it only suffices to make sure there is a gain mechanism before bringing pump phase noise in the picture. 

\section{Phase Noise in Four-wave mixing (4WM) and Three-wave mixing (3WM) processes}
Now we assume that the microwave pump tone is generated by an oscillator with phase noise. An oscillator with phase noise is modeled using \textbf{OSCwPhNoise} in Keysight ADS simulator. The phase-noise power spectrum is defined by the amount of power at each offset frequency around the main pump tone to simulate the Leeson model \cite{Leeson1966, Maas2005}. This is a semi-empirical model which accounts for several sources of noise that manifest themselves as the phase fluctuations of the oscillator. The sources are for example thermal and 1/f (flicker) noise on the control voltage lines in a voltage-controlled oscillator (VCO). The mathematical form of phase noise skirt or spectrum density around the main pump tone is:

\begin{equation}
\label{eqn:Leeson}
S_{\phi}=\frac{Fk_BT}{2P_{tone}}\left[1+(\frac{f_{center}}{2Q_L\Delta f})^2\right](1+\frac{f_{corner}}{\Delta f}),
\end{equation}

\noindent where $F$ is an empirical factor, $k_B$, $T$, and $Q_L$ are Boltzmann constant, absolute temperature, and loaded quality factor of the LC oscillator, respectively. $f_{center}$ is the oscillator's main tone (here the pump), $\Delta f$ is the frequency offset away from the main tone and $f_{corner}$ is the frequency offset at which the slope of the noise skirt around the main tone changes from $1/f^3$ to $1/f^2$ or from $1/f^2$ to $1/f$. This depends on the exact sources of the phase noise (thermal versus flicker). Studying all mechanisms which lead to phase-noise in an oscillator is beyond the scope of this paper, however it suffices to know that the thermal and flicker noise on supply and control lines of oscillators generally lead to phase noise. Also through the process of AM to PM conversion some or all power of amplitude noise can be converted to phase noise due to nonlinearity of devices making the oscillator, mixers or amplifiers in the output chain of a signal generator \cite{Gardner2005, Leeson1966}.
Alternative to Leeson model, there is a user-defined method of giving the values of noise spectral density at as many offset frequency points as desired \textit{e.g.,} by defining a list like \textbf{PhaseNoise=list(10Hz,-50dB, 100Hz,-70dB, 1kHz,-90dB, ...)}, where the unit of power spectrum is dBc/Hz (decibel with respect to the main tone or the carrier). This helps us to simulate the effect of phase noise without being concerned about the exact source of it in the off-the-shelf oscillator used as the pump source. \\ 

\begin{figure}[t!] 
\centering
\includegraphics[width=3.5in]{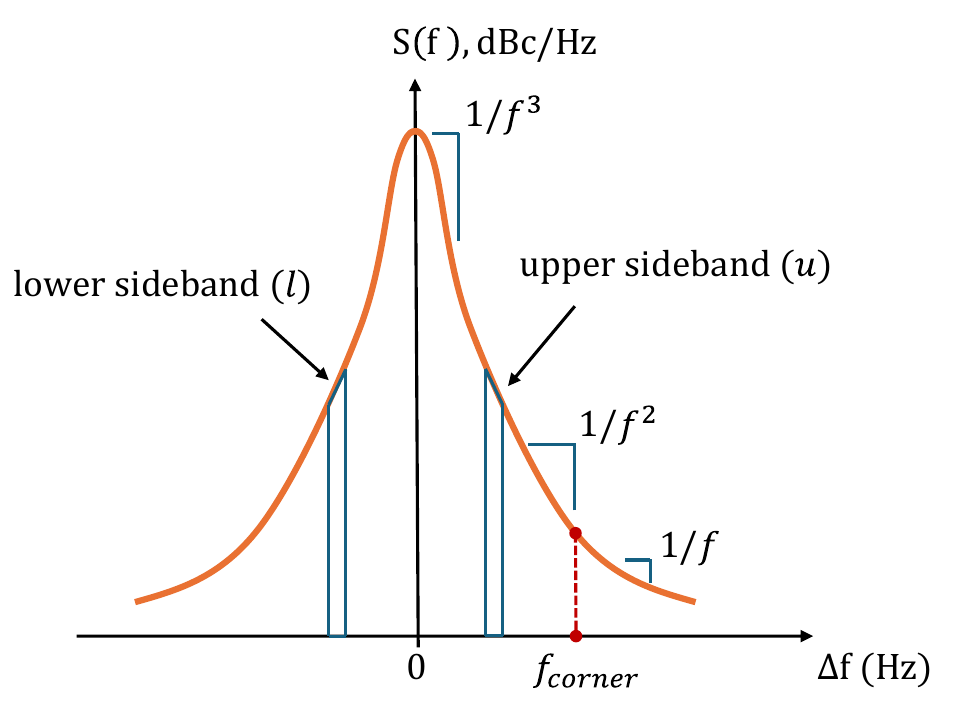}
\caption{Double side-band (DSB) phase noise power spectrum of a typical oscillator with phase noise fitted to Leeson model. Note that single side band (SSB) spectrum is the folded version of DSB. The number of corner frequencies depends on the type pf the original noise which causes the phase noise. The width of the blue trapezoid shows the integration bandwidth in calculating the total noise power e.g., 1~Hz. }
\label{PNoise_skirt}
\end{figure} 

\begin{figure}[t]
    \centering
    \includegraphics[width=\columnwidth]{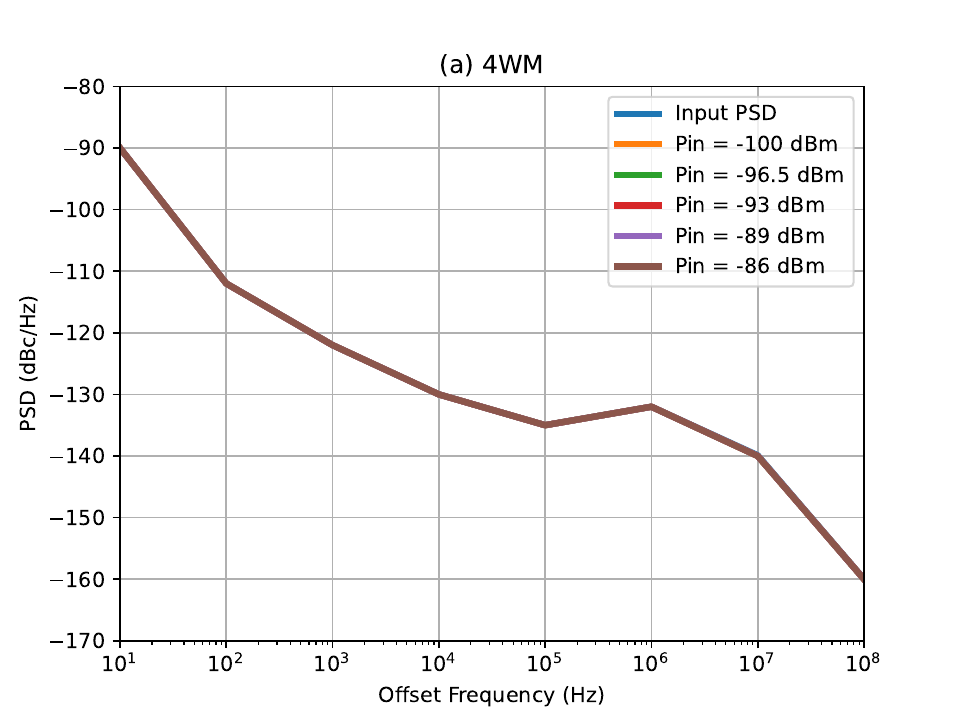}
    \label{fig:first}
    \vspace{1mm}
    \includegraphics[width=\columnwidth]{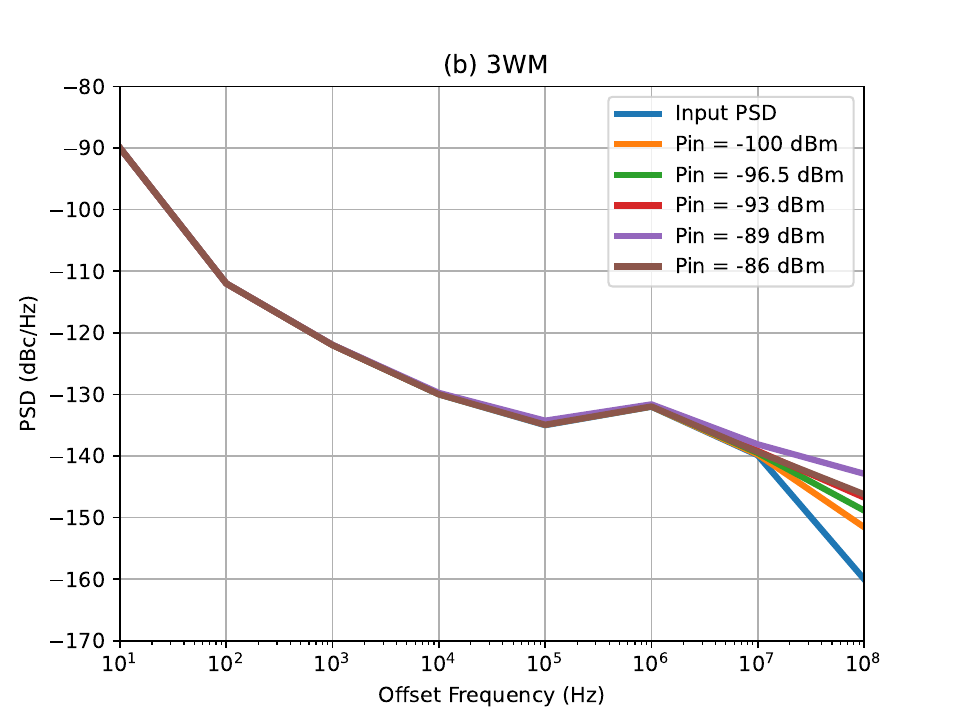}
    \label{fig:second}
    \caption{ (a) SSB phase noise spectra at the input and output nodes of a 4WM JTWPA. The horizontal axis is the offset frequency from the pump frequency. The pump power is swept from -100~dBm to -86~dBm. For all sweep values, the output spectrum is the same as the input one. (b) The SSB phase noise spectra at the output node of 3WM JTWPA. The input noise spectrum is the same as for (a). In both simulations, the output phase noise data of AnaPico AP5022A signal generator is used.}
    \label{Anapico_sim}
\end{figure}

As input pump phase noise, we assume that the noise spectrum represents that of AP5022A AnaPico signal generator \cite{AnaPcio}. The phase noise spectrum of this model at 10~GHz is defined by the list (one point per frequency decade) following its datasheet. Note that modern RF sources usually are equipped with digitally-programmed frequency synthesizers with phase locked loop (PLL). Hence, by virtue of the inherent low-pass filtering of feedback in the PLL, the low frequency content of phase noise spectrum (immediately after and before the main tone) is significantly reduced. \\
\indent With the given input phase noise spectrum, the HB noise analysis program calculates the spectral density for both phase noise ($S_\phi$) and amplitude noise ($S_a$) at the output or any node of the JTWPA. For this, HB uses the values of upper and lower side-bands [See Fig.\ref{PNoise_skirt}] of noise as follows:

\begin{eqnarray}
\label{eqn:spectral}
S_{\phi}=\frac{<v_u,v_u^*>+<v_l,v_l^*>-2\text{Re}[<v_u,v_l^*>e^{2j\phi}]}{2V_{p}^2}, \label{sub:1}\\
S_{a}=\frac{<v_u,v_u^*>+<v_l,v_l^*>+2\text{Re}[<v_u,v_l^*>e^{2j\phi}]}{2V_{p}^2},\label{sub:2}
\end{eqnarray}

\noindent where $V_{p}$ is the amplitude of the main pump tone and the terms like $<v_u,v_l^*>$ are cross-correlation of upper ($u$) and lower ($l$) side-band noise components. They are off-diagonal elements of the voltage noise correlation matrix defined in \cite{Maas2005}. The phase difference between the upper and lower side-band components is $\phi$. The SSB phase noise spectrum is then found from Eq. \ref{sub:1} by $\mathcal{L}=S_\phi/2$ \cite{Maas2005}.
The spectral densities $S_{\phi}$ and $S_{a}$ in Eqs. \ref{sub:1} and \ref{sub:2} are accessible in Keysight ADS simulator as \textbf{NodeName.pnmx} and \textbf{NodeName.anmx}, respectively, where \textbf{NodeName} is the name of the circuit node. \\
As observed in Fig. \ref{Anapico_sim}(a), the noise spectra of both input and output in 4WM JTWPA are equal and there is no sign of increase or decrease of phase noise at the output of the JTWPA regardless of the pump power value. The pump power is swept from -100~dBm to -86~dBm. Note that higher power brings JJs closer to their critical point (superconducting to normal transition), rendering the convergence of HB analysis difficult. This threshold which is around -73~dBm, was therefore avoided in the simulation. Since the spectrum of phase noise in the AnaPico model is low-pass filtered, the shape does not follow Leeson formula, but \textbf{PhaseNoise} list in ADS allows the piecewise definition of any phase noise spectrum. 

In contrast to 4WM JTWPA, the power density increases at higher frequency offsets in 3WM JTWPA as shown in Fig. \ref{Anapico_sim}(b). For example, at 100~MHz offset from the pump tone, the spectral density grows at the output node of 3WM JTWPA by about 20~dB when pump power is increased from -100~dBm to -86~dBm. The input noise spectrum or phase-noise skirt is the same as the one used for 4WM case in Fig. \ref{Anapico_sim}(a). The phase noise skirt increases incrementally at high frequency offsets along the length of the 3WM JTWPA. For example, in the middle of the TWPA \textit{e.g.,} at 500'th unit cell, there is 10~dB increase at 100~MHz offset at -86~dBm of pump power.

Answering this question is now called for: Why there is an increase of phase noise spectral density in 3WM JTWPA but not in the 4WM JTWPA? 

\section{A nonlinear transfer function model}
\begin{figure}[t!] 
\centering
\includegraphics[width=3.5in]{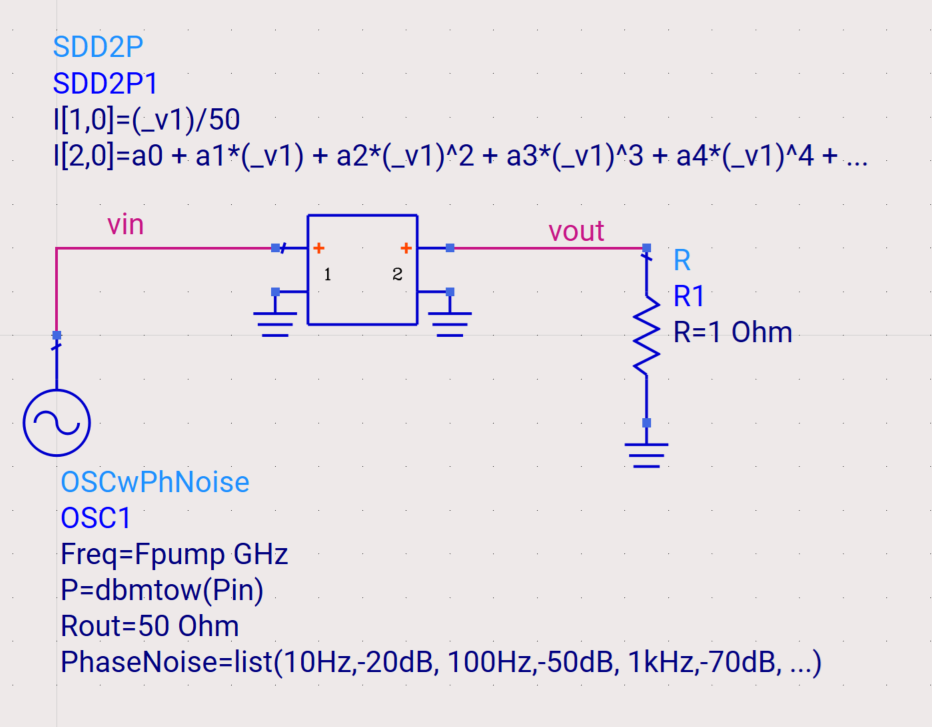}
\caption{The nonlinear transfer function implemented with SDD in Keysight ADS simulator for understanding the effect of odd and even nonlinearities on the output phase noise.}
\label{ideal_SDD}
\end{figure}

\begin{figure}[t]
    \centering
    \includegraphics[width=\columnwidth]{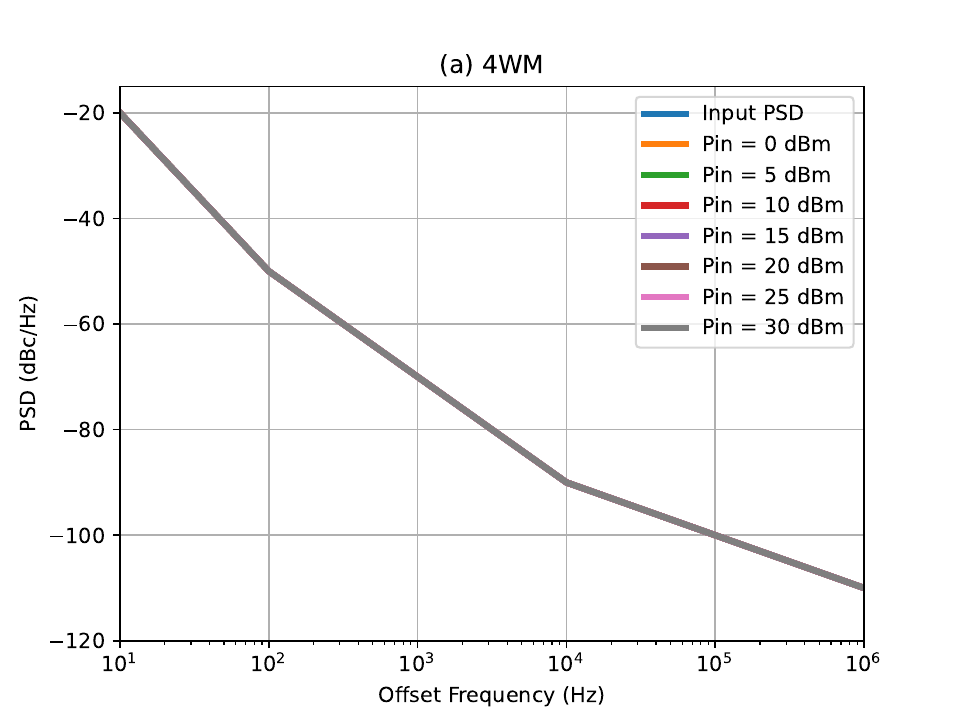}
    \vspace{1mm}
    \includegraphics[width=\columnwidth]{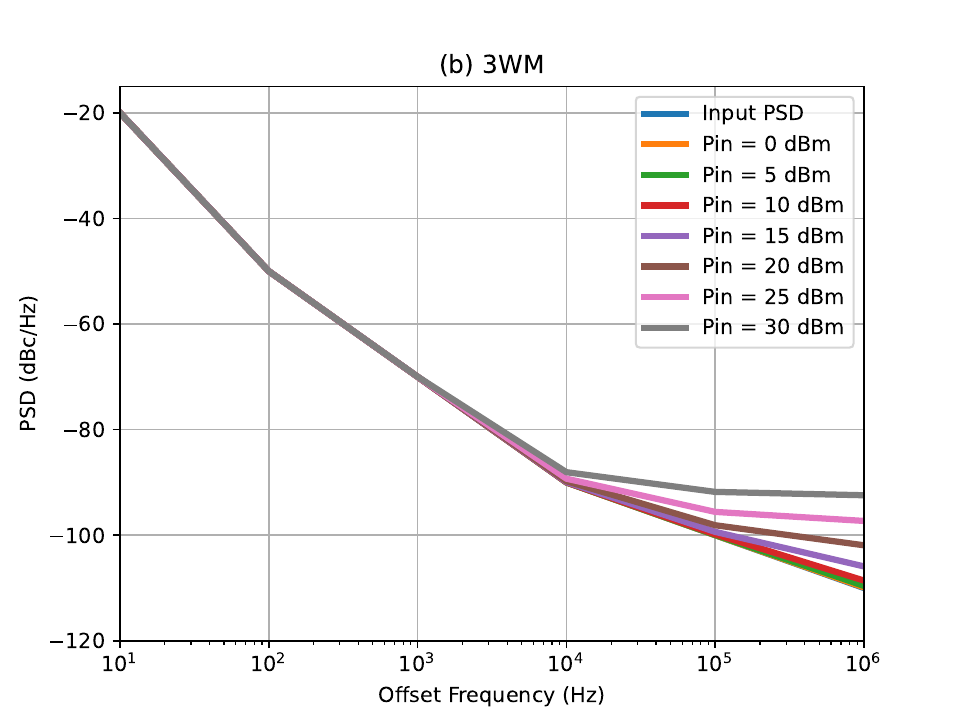}
    \caption{Effect of even order nonlinearities at the output of the model amplifier. (a) When the even order terms are zero \textit{i.e.,} $a_0=a_2=a_4=0$, (b) Adding the even order terms ($a_0=a_2=a_4=1$) recreates the same results obtained for 3WM JTWPA.}
    \label{SDDmodel_1MHz}
\end{figure}

To answer the above question, we consider a nonlinear amplifier modeled by a polynomial as the transfer function \textit{i.e.,} $y = a_0 + a_1x + a_2x^2 + a_3x^3 +a_4x^4$, where $x$ and $y$ are input and output signals, and $a_1$ can be considered as small-signal linear gain, respectively. We repeat HB phase noise analysis simulations to obtain a deeper understanding of the mechanisms responsible for the observations in section II. Fig. \ref{ideal_SDD} shows the nonlinear transfer function model implemented with symbolically-defined devices (SDD) in Keysight ADS simulator. The phase noise simulation shows that depending on the values of odd or even order nonlinear terms, the phase noise spectrum shows insignificant or considerable change with the increase of pump power. If all even coefficients are set to zero ($a_0=a_2=a_4=0$), there is no change in the spectral density of phase noise as can be seen in Fig. \ref{SDDmodel_1MHz}(a). This is what we already observed in 
Fig. \ref{Anapico_sim}(a) for 4WM JTWPA where the output spectrum contains odd harmonics of the pump. 

When the even order nonlinearities are turned on \textit{i.e.,} $a_0=a_2=a_4=1$, we see the same observation we had in Fig. \ref{Anapico_sim}(b) \textit{i.e.,} the increase of spectral density as the pump power ramps up. Fig. \ref{SDDmodel_1MHz}(b) suggests that indeed in 3WM process, even-order nonlinearity terms are responsible for the spectral growth at higher offsets around the pump tone when the power of pump is increased. 
Note that in the ideal nonlinear amplifier model, the polynomial coefficients are normalized between 0 and 1 and the power values in this model and Fig. \ref{SDDmodel_1MHz} do not need to correspond to the powers of real JTWPA and a real signal source data for simulations in Section II.

\section{Cyclo-stationary noise (periodic correlation)}
\begin{figure}[t!] 
\centering
\includegraphics[width=3.5in]{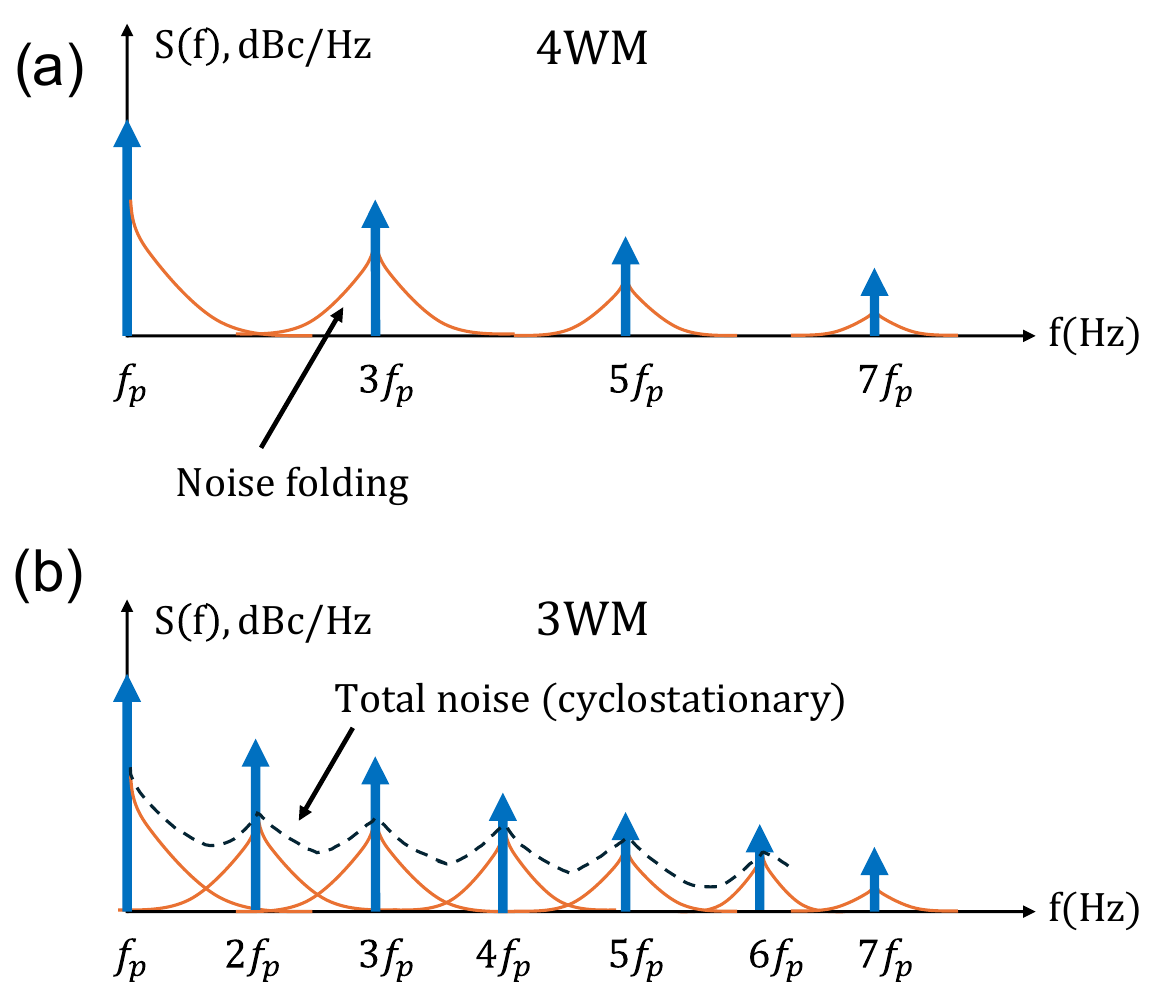}
\caption{Emergence of cyclo-stationary noise (and periodic correlation) in frequency domain due to nonlinearity and periodic modulation of circuit operating point under the influence of the large pump.}
\label{cyclo-stat}
\end{figure}

The previous observations all hint at cyclo-stationary property of noise in nonlinear periodic circuits \cite{Kundert1999}. Applying  a large pump to the nonlinear circuits leads to the periodic change of the operating point around which noise and signal are applied. As a result of this periodic modulation of noise in time domain, the noise spectrum is not flat and there are periodic replicas and correlation between each side band around the harmonics of the pump. As it is shown in Fig. \ref{cyclo-stat}(a), in 4WM JTWPA the phase noise spectra are shifted around the odd harmonics of the pump which are far from each other and the overlapping of skirts are not significant, hence correlations between side bands are small, even by increasing the power of the pump (or height of each harmonic).

\begin{figure}[t]
    \centering
    \includegraphics[width=\columnwidth]{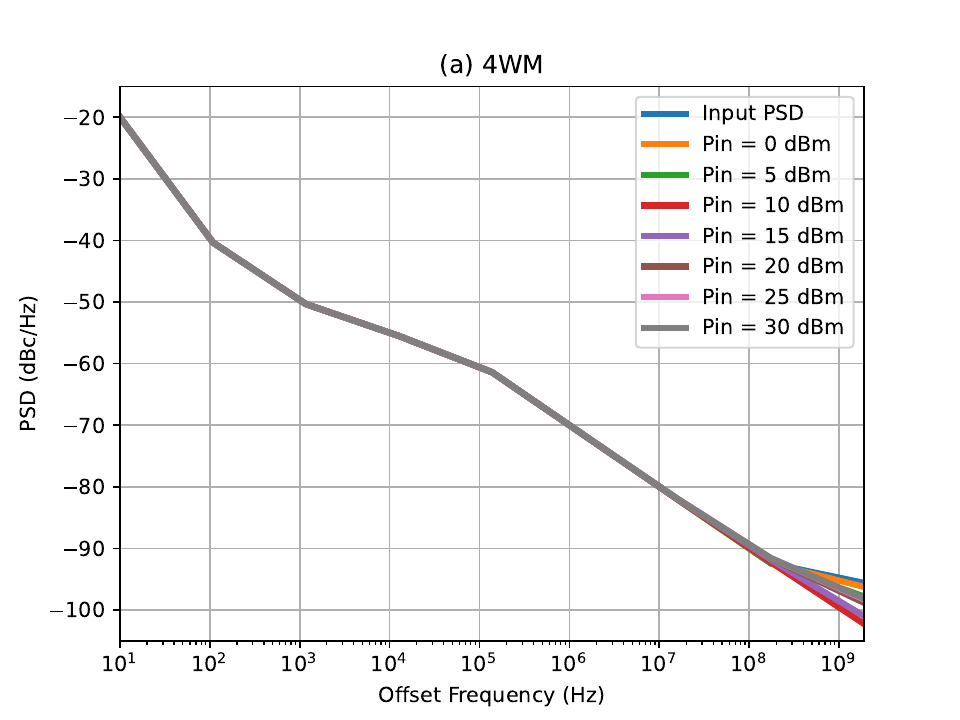}
    \vspace{1mm}
    \includegraphics[width=\columnwidth]{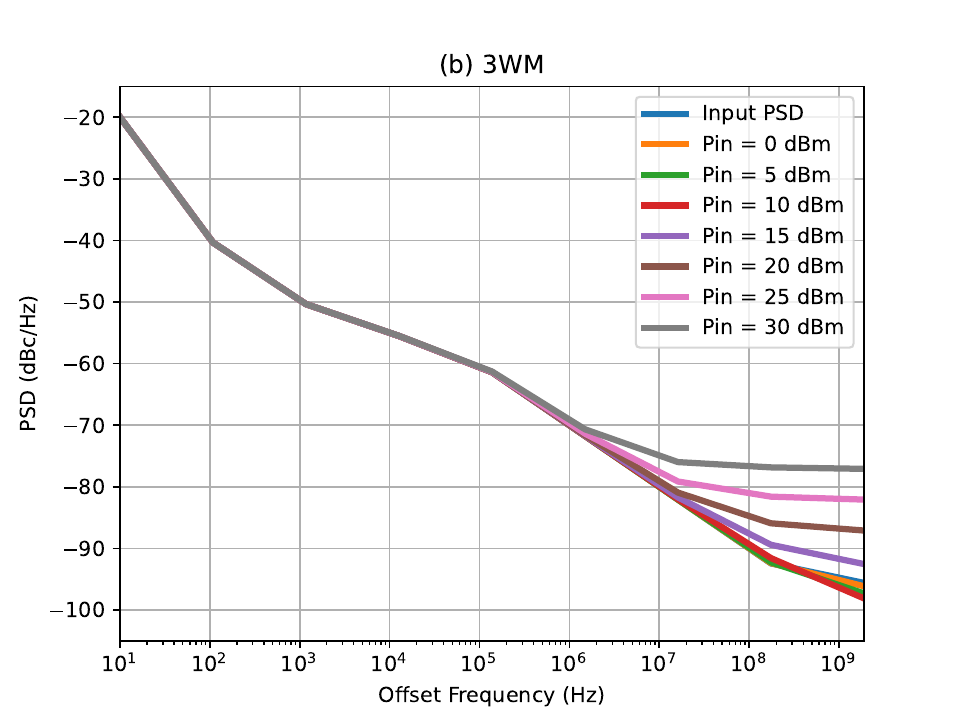}
    \caption{(a) Test of nonlinear amplifier output phase noise with large frequency offset of $\Delta f = 1.9~\text{GHz}$ and a lower pump frequency of $f_p=1~\text{GHz}$. Similar to Fig. \ref{SDDmodel_1MHz}(a), only odd harmonics of the pump are included ($a_0=a_2=a_4=0$). (b) The same results as above when all harmonics of $f_p=1~\text{GHz}$ are included \textit{i.e.,} $a_0=a_2=a_4=1$.}
    \label{SDDmodel_1p9ghz}
\end{figure}

In contract to 4WM case, in 3WM JTWPA, the phase noise spectrum is shifted and centered around all harmonics of the pump which causes more overlapping of skirts [See Fig. \ref{cyclo-stat}(b)]. Increasing the pump power here leads to more overlapping and correlation due to closeness of the harmonics, in line with our observations in both HB simulations of JTWPA and the simple polynomial model.
To further test the theory of cyclo-stationary noise, we intentionally decreased the pump frequency to 1~GHz and run the HB periodic noise analysis from 1~Hz to 1.9~GHz offset around the main tone (Note that HB analysis does not allow the offset coincide with main tone harmonics \textit{e.g.,} 2~GHz). Figure \ref{SDDmodel_1p9ghz}(a) shows that for odd only (4WM) process, the overlapping of noise spectra is not significant as the closest noise skirt is coming from $3f_p = 3~\text{GHz}$. On the other hand, for all pump harmonics (3WM), the overlapping due to the noise skirt around $2f_p=2~\text{GHz}$ is clearly seen in Fig. \ref{SDDmodel_1p9ghz}(b) as well as the effect of pump power in increasing it. 


\section{Conclusions}
This study suggests that the phase noise spectrum regrowth at the output of JTWPA is more significant for 3WM type when the pump power increases. Both 4WM and 3WM JTWPAs were studied using harmonic balance noise analysis in Keysight ADS simulator. The difference between 3WM and 4WM phase noise behavior at high pump powers was explained using a polynomial-based nonlinear amplifier model and the concept of cyclo-stationary noise. This model agrees with the HB noise analysis results of JTWPA. Even harmonics (in 3WM process) are what cause more correlations between phase noise subbands and increase of noise power at higher frequency offsets. 
Since 3WM JTWPAs are favorite because of their promised high gain, high bandwidth, and importantly, out of the qubit band pump frequency, care must be taken in how much output phase noise can be tolerated in these amplifiers. This also mandates more study and better quantification of phase noise on JTWPA figures of merit \textit{e.g.,} quantum efficiency \cite{peng2022x}, SNR, and gain. 
For the case of microwave pulses which control the qubits \textit{i.e.,} change the state or frequency of the qubits (XY and flux pulses, respectively), the effect of master clock LO phase noise was analyzed in \cite{Ball2016}. This effect is quantified by amount of dephasing of qubits in this context. Note that the input phase noise which we use in HB noise analysis is the real output data of AnaPico signal generator with advanced phase noise reduction around the main tone. Using other sources with higher phase noise power around the main tone (or sources lacking PLL or advanced filtering) may lead to higher output phase noise regrowth than what we calculated in this study.   

\bibliographystyle{IEEEtran}
\bibliography{IEEEabrv,Maindraft_IEEEQW_Canada2026_DShirietal}

@article{Sandbo2025,
  title = {Josephson traveling-wave parametric amplifier based on a low-intrinsic-loss lumped-element coplanar waveguide},
  author = {Chang, C.W. Sandbo and Van Loo, Arjan F. and Hung, Chih-Chiao and Zhou, Yu and Gnandt, Christian and Tamate, Shuhei and Nakamura, Yasunobu},
  journal = {Phys. Rev. Appl.},
  volume = {24},
  issue = {4},
  pages = {044081},
  numpages = {24},
  year = {2025},
  month = {Oct},
  publisher = {American Physical Society},
  doi = {10.1103/qhl6-cz2z},
}

@article{Armstrong1962,
  title = {Interactions between Light Waves in a Nonlinear Dielectric},
  author = {Armstrong, J. A. and Bloembergen, N. and Ducuing, J. and Pershan, P. S.},
  journal = {Phys. Rev.},
  volume = {127},
  issue = {6},
  pages = {1918--1939},
  numpages = {0},
  year = {1962},
  month = {Sep},
  publisher = {American Physical Society},
  doi = {10.1103/PhysRev.127.1918},
}

@misc{AnaPcio,
      title={{AP}5022{A} datasheet, {K}eysight {E}xpert {A}nalog {S}ignal {G}enerators}, 
      author={},
      url = {https://www.keysight.com/us/en/product/AP5022A/analog-signal-generator-9-khz-to-54-ghz.html},
}

@article{Josephson1962,
title = {Possible new effects in superconductive tunnelling},
journal = {Physics Letters},
volume = {1},
number = {7},
pages = {251-253},
year = {1962},
issn = {0031-9163},
author = {B.D. Josephson}
}

@book{Gardner2005,
  title={Phaselock Techniques},
  author={Floyd M. Gardner},
  year={2005},
  month ={July},
  publisher={John Wiley and Sons, Inc.},}

@ARTICLE{Kundert1999,
  author={Kundert, K.S.},
  journal={IEEE Journal of Solid-State Circuits}, 
  title={Introduction to {RF} simulation and its application}, 
  year={1999},
  volume={34},
  number={9},
  pages={1298-1319},
  doi={10.1109/4.782091}}

@ARTICLE{Leeson1966,
  author={Leeson, D.B.},
  journal={Proceedings of the IEEE}, 
  title={A simple model of feedback oscillator noise spectrum}, 
  year={1966},
  volume={54},
  number={2},
  pages={329-330},
  doi={10.1109/PROC.1966.4682}}

@article{Ball2016,
  author    = {Harrison Ball and William D. Oliver and Michael J. Biercuk},
  title     = {The role of master clock stability in quantum information processing},
  journal   = {npj Quantum Information},
  volume    = {2},
  number    = {1},
  pages     = {16033},
  year      = {2016},
  doi       = {10.1038/npjqi.2016.33},
}

@INPROCEEDINGS{Likai2025,
  author={Yang, Likai and Wang, Jennifer and Hassan, Mohamed Awida and Krantz, Philip and O'Brien, Kevin P.},
  booktitle={2025 IEEE/MTT-S International Microwave Symposium - IMS 2025}, 
  title={Modeling {J}osephson Traveling-Wave Parametric Amplifiers with Electromagnetic and Circuit Co-Simulation}, 
  year={2025},
  volume={},
  number={},
  pages={65-68},
  doi={10.1109/IMS40360.2025.11103819}}

@book{Maas2005,
  title={Noise in Linear and Nonlinear Circuits},
  author={Setphen Maas},
  year={2005},
  month ={August},
  publisher={Artech House Publishers},}

@article{Fadavi2023,
    author = {Fadavi Roudsari, Anita and Shiri, Daryoush and Renberg Nilsson, Hampus and Tancredi, Giovanna and Osman, Amr and Svensson, Ida-Maria and Kudra, Marina and Rommel, Marcus and Bylander, Jonas and Shumeiko, Vitaly and Delsing, Per},
    title = {Three-wave mixing traveling-wave parametric amplifier with periodic variation of the circuit parameters},
    journal = {Applied Physics Letters},
    volume = {122},
    number = {5},
    pages = {052601},
    year = {2023},
    month = {Jan},
    issn = {0003-6951},
    doi = {10.1063/5.0127690},
}

@article{Landauer1960,
    author = {Landauer, Rolf},
    title = {Parametric amplification along nonlinear transmission lines},
    journal = {Journal of Applied Physics},
    volume = {31},
    number = {3},
    pages = {479-484},
    year = {1960},
    month = {Mar},
    issn = {0021-8979},
    doi = {10.1063/1.1735612},
}

@article{krantz2019quantum,
	author = {Krantz, P. and Kjaergaard, M. and Yan, F. and Orlando, T. P. and Gustavsson, S. and Oliver, W. D.},
	journal = {Applied Physics Reviews},
	month = {Jun},
	number = {2},
	title = {{A quantum engineer's guide to superconducting qubits}},
	volume = {6},
	year = {2019},
	doi = {10.1063/1.5089550}
}

@article{macklin2015near,
	author = {Macklin, C. and O’Brien, K. and Hover, D. and Schwartz, M. E. and Bolkhovsky, V. and Zhang, X. and Oliver, W. D. and Siddiqi, I.},
	journal = {Science},
	month = {Sept},
	number = {6258},
	pages = {307--310},
	title = {{A near–quantum-limited Josephson traveling-wave parametric amplifier}},
	volume = {350},
	year = {2015},
	doi = {10.1126/science.aaa8525}
}

@article{malnou2021three,
	author = {Malnou, M. and Vissers and Wheeler, J.D. and Aumentado, J. and Hubmayr, J. and Ullom, J.N. and Gao, J.},
	journal = {PRX Quantum},
	month = {Jan},
	number = {1},
	title = {{Three-wave mixing kinetic inductance traveling-wave amplifier with near-quantum-limited noise performance}},
	volume = {2},
	year = {2021},
	doi = {10.1103/prxquantum.2.010302}
}

@article{peng2022x,
	author = {Peng, Kaidong and Poore, Rick and Krantz, Philip and Root, David E. and O'Brien, Kevin P.},
	journal = {2022 IEEE International Conference on Quantum Computing and Engineering (QCE)},
	month = {Sept},
	pages = {331--340},
	title = {X-parameter based design and simulation of {J}osephson traveling-wave parametric amplifiers for quantum computing applications},
	year = {2022},
	doi = {10.1109/qce53715.2022.00054}
}

@article{aumentado2020superconducting,
	author = {Aumentado, Jose},
	journal = {IEEE Microwave Magazine},
	month = {July},
	number = {8},
	pages = {45--59},
	title = {{Superconducting parametric amplifiers: The state of the art in Josephson parametric amplifiers}},
	volume = {21},
	year = {2020},
	doi = {10.1109/mmm.2020.2993476}
}

@article{shiri2024modeling,
	author = {Shiri, Daryoush and Nilsson, Hampus Renberg and Telluri, Pavan and Roudsari, Anita Fadavi and Shumeiko, Vitaly and Fager, Christian and Delsing, Per},
	journal = {IEEE Microwave Magazine},
	month = {Oct},
	number = {11},
	pages = {54--73},
	title = {{Modeling and harmonic balance analysis of superconducting parametric amplifiers for qubit readout: a tutorial}},
	volume = {25},
	year = {2024},
	doi = {10.1109/mmm.2024.3429141}
}

@phdthesis{HamnpusThesis,
  author       = {Hampus Renberg Nilsson},
  title        = {Superconducting lumped-element travelling-wave parametric amplifiers},
  school       = {Chalmers University of Technology},
  year         = {2024},
  type         = {Ph.D. thesis},
  address      = {Gothenburg, Sweden},
  month        = {September},
}

@ARTICLE{Mohebbi09,
  author={Mohebbi, Hamid Reza and Majedi, A. Hamed},
  journal={IEEE Transactions on Applied Superconductivity}, 
  title={Periodic Superconducting Microstrip Line With Nonlinear Kinetic Inductance}, 
  year={2009},
  volume={19},
  number={3},
  pages={930-935},
  doi={10.1109/TASC.2009.2018085}
}

@article{Peng22floquet,
  title = {Floquet-Mode Traveling-Wave Parametric Amplifiers},
  author = {Peng, Kaidong and Naghiloo, Mahdi and Wang, Jennifer and Cunningham, Gregory D. and Ye, Yufeng and O'Brien, Kevin P.},
  journal = {PRX Quantum},
  volume = {3},
  issue = {2},
  pages = {020306},
  numpages = {20},
  year = {2022},
  month = {Apr},
  publisher = {American Physical Society},
  doi = {10.1103/PRXQuantum.3.020306}
}
\end{document}